\documentclass[11pt,a4paper,fleqn]{article}
\usepackage[russian,english]{babel}
\usepackage{epsfig}
\usepackage[cp1251]{inputenc}
\usepackage{floatrow}
\usepackage{amssymb,amsmath,amsfonts,amsthm,graphicx,psfrag}
\usepackage{indentfirst}
\usepackage{hyperref}
\usepackage[title,titletoc]{appendix}
\usepackage{graphicx}
\usepackage{amsfonts}
\usepackage{bm}
\usepackage{verbatim}
\usepackage{epsfig}
\graphicspath{{./pictures/}}
\usepackage{authblk}
\usepackage{cite}
\usepackage[title,titletoc]{appendix}
\usepackage{slashed}
\usepackage{amsmath}
\usepackage[left=2cm,right=2cm,
    top=2cm,bottom=2cm,bindingoffset=0cm]{geometry}
		\newcommand{\be}{\begin{eqnarray}}
\newcommand{\ee}{\end{eqnarray}}
\newcommand{\ba}{\begin{aligned}}
\newcommand{\ea}{\end{aligned}}

\newcommand{\lan}{\langle}
\newcommand{\ran}{\rangle}

\usepackage{authblk}

\title{Colour-electric and colour-magnetic confinement}
\author[*]{N.O.~Agasian}
\author[*,+]{Z.V.~Khaidukov}
\author[*]{M.S.~Lukashov}
\author[*]{Yu.A.~Simonov}

\affil[*]{NRC ``Kurchatov Institute'', Moscow 117259, Russia}
\affil[+]{Moscow Institute of Physics and Technology,
Dolgoprudny 141700, Moscow Region, Russia}

\begin{document}

\maketitle
\begin{abstract}
The basic properties of the confinement mechanism in QCD -- the temperature dependence of the spatial and temporal string tensions ($\sigma_s(T)$ and $\sigma_E(T)$) -- are studied in the framework of the Field Correlator Method  (FCM). It is shown that both functions are connected respectively to the spatial and temporal parts of the vacuum gluon energy $\epsilon_s$ and $\epsilon_E$ which define their equal values at $T=0$. However at $T>0$ the  spatial part is growing with $T$ while the temporal part is destroyed by the hadronic pressure at $T=T_c$ (the deconfinement). Both properties are derived within the same method and are in a good agreement with the corresponding  lattice data.
\end{abstract}

\label{sec:intro}

The confinement in QCD is the most intriguing phenomenon which ensures more than 90 percent of the visible mass in the Universe and makes the world around us such as we see it. We  can investigate confinement  with the help of   the Field Correlator Method (FCM) \cite{1,2,3,4,5,5*}, this is the  theory in which the ``Area law'' between colour charges  at zero temperature exists. Having such a basis, we can try to study the temperature dependence of QCD.  
To describe vacuum field correlators we need to decompose gluon field on  the
colorelectric (CE) and the colormagnetic (CM) parts $E_i^a,H_i^a$. At zero temperature
$T=0$ the behaviour of all physical quantities is expressed  via the basic nonperturbative
parameter -- the string tension, which  can have different values in the light-like  $\sigma_E$ and space-like $\sigma_H$ areas, but  at  zero temperature  $\sigma_E(T=0)=\sigma_H(T=0)= \sigma$.  Very important role in FCM plays
bilocal correlator (BC) of gluonic fields strength:
	\be
	\frac{g^{2}}{N_{c}}<tr_{f}\Phi(y,x)F_{\mu\nu}(x)\Phi(x,y)F_{\lambda \rho}(y)>\equiv D_{\mu\nu,\lambda\rho}(x,y) \label{eqcorr}.
	\ee
From this moment we use     $F_{\mu\nu}\equiv F^{a}_{\mu\nu}T^{a}$, $a=1..N_{f},T^{a}$ - are generators of fundamental representation of  $SU(N_{c})$
. In Eq. (\ref{eqcorr}) symbol $<>$  means  averaging over Yang-Mills action  $S=\frac{1}{4g^{2}}\int{d^{4}x}(F^{a}_{\mu\nu}), F^{a}_{\mu\nu}=\partial_{\mu}A_{\nu}-\partial_{\nu}A_{\mu}+gf^{abc}A^{b}_{\mu}A^{c}_{\nu}, a=1..N^{2}_{c}-1$, $\Phi(x,y)=Pexp(i\int^{x}_{y}A_{\mu}dz^{\mu}), \mu=1..4$\footnote{We work in Euclidean space, the fourth component plays role of Euclidean time.} is  the Wilson line in fundamental representation.
	 We  can write BC as follows:
	\be
	 D_{\mu\nu,\lambda\rho}(x,y)=(\delta_{\mu\lambda}\delta_{\nu\rho}-\delta_{\mu\rho}\delta_{\nu\lambda})D(x-y)+\frac{1}{2}(\frac{\partial}{\partial x_{\mu}}(x-y)_{\lambda}\delta_{\nu\rho}+perm.))D_{1}(x-y)  \label{eqcorr1},
	\ee
	$D(x-y),D_{1}(x-y)$ - are scalar functions.  We also can add index E or H\footnote{We will write them only where we need to avoid ambiguity.} to  $D,D_{1}$, because $E_{i}=F_{0i},H_{i}=\epsilon_{ijk}F^{jk}/2 , i=1,2,3$ and $<EH>=0$.
	Functions  $D^{E,H}(x),D^{E,H}_{1}(x)$   define all confining QCD dynamics and in particular the string tensions:
\be
\sigma_E= 1/2 \int (d^2z)_{i4} D^{E}(z), \sigma_H= 1/2 \int (d^2z)_{ik} D^{H}(z) \label{2}. \ee
 The most interesting fact is  that
 at $T>0$  $\sigma_E(T)$ and $\sigma_H(T)= \sigma_s(T)$
behave differently. Namely:  $\sigma_E(T)$ displays a spectacular drop before $T=T_c$ and
disappears above $T=T_c$, while in contrast to that $\sigma_s(T)$
grows almost quadratically at large $T$, as  was
found on the lattice \cite{4,7*,8,9} and supported by the studies
in the framework of the FCM \cite{10,10*,11,12,12*,12**,18,15,14}.  At next sections we will discuss dereferences in  behaviour of the colour-magnetic and colour electric string tensions, and try to make some predictions about their temperature dependence.

\section{Colour-magnetic string tension at non-zero T}
In this section we will focus on the colour-magnetic string tension (CMST)  in gluodynamics. We can write   BC as:
\be
D_{\mu\nu,\lambda\rho}(x,y)=D^{0}_{\mu\nu,\lambda\rho}(x,y)+D^{1}_{\mu\nu,\lambda\rho}(x,y)+D^{2}_{\mu\nu,\lambda\rho}(x,y),
\ee
where the number at the top of the letter D means power minus two of coupling constant g.  For $D^{0}(x,y)$ we obtain:
\be
D^{0}_{\mu\nu,\lambda\rho}(x,y)=\frac{g^{2}}{2N_{c}^{2}}( \frac{\partial}{\partial x_{\mu}} \frac{\partial}{\partial y_{\nu}} G^{1g}(x,y) +perm. )+\Delta^{0}_{\mu\nu,\lambda\rho},
\ee
where $\Delta^{0}_{\mu\nu,\lambda\rho}(x,y)$ contains contribution of higher field cumulants, which we systematically
discard. Here  $G^{1g}$ is one-gluelump Green's function:
\be
G^{1g}_{\mu\nu}(x,y)=<tr_{adj}\hat{A}_{\mu}(x)\hat{\Phi}_{adj}(x,y)\hat{A}_{\nu}(y)>, \label{1Gl}
\ee
 $tr_{adj}$ is a trace  over adjoint indices.
The expression for $D^{2}_{\mu\nu,\lambda\rho}(x,y)$ reads as:
\be
D^{2}_{\mu\nu,\lambda\rho}(x,y)=-\frac{g^{4}}{2N^{2}_{c}}<tr_{adj}([A_{\mu}(x),A_{\nu}(x)]\hat{\Phi}(x,y)[A_{\lambda}(x),A_{\rho}(x)])> \label{35}.
\ee
We remind  that:
\be
[A_{i},A_{k}]=iA^{a}_{i}A^{b}_{k}f^{abc}T^{c}.
\ee
	Let's consider:
\be
G_{\mu\nu,\lambda\rho}(x,y)=tr_{adj}<f^{abc}f^{def}A^{a}_{\mu}(x)A^{b}_{\nu}(x)T^{c}\hat{\Phi}(x,y)A^{d}_{\lambda}(y)A^{e}_{\rho}(y)T^{f}>.\label{eqgluelump}
\ee
We can represent $G_{\mu\nu,\lambda\rho}(x,y)$ in
the form:
\be
G_{\mu\nu,\lambda\rho}(x,y)=N^{2}_{c}(N^{2}_{c}-1)(\delta_{\mu\lambda}\delta_{\nu\rho}-\delta_{\mu\rho}\delta_{\nu\lambda})G^{2gl}(x,y), \label{38}
\ee
where $G^{2gl}(x,y)$ is the Green's function of the two-gluon gluelump.\par
Comparison of Eqs. (\ref{eqcorr1}), (\ref{35}) and (\ref{38}) immediately yields the following
expression for D(x-y):
\be
D(x-y)=\frac{g^{4}(N^{2}_{c}-1)}{2}G^{2gl}(x,y) \label{gl}.
\ee
Thus if one calculate two-gluon gluelump Greens function at non-zero temperature he will obtain CMST.  That have been done in \cite{SAK}. For CMST  Fig.\ref{FiG2} have been obtained:

\begin{figure}
\center{\includegraphics[width=0.9\linewidth]{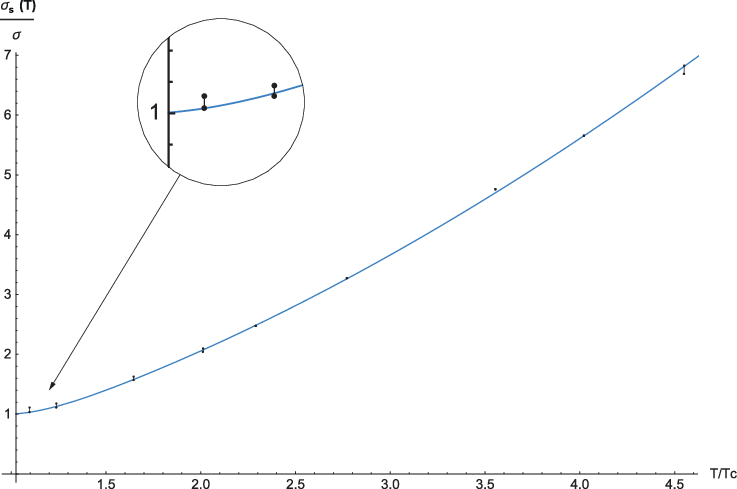}}
    \caption{Spatial string tension $\sigma_s (T)/\sigma$ for SU(3)
gauge theory as function of $T/T_c$.  The lattice data with errors are from Ref.\cite{8}. $T_{c}$=270 MeV}
    \label{FiG2}
\end{figure}

One can see that according to FCM at temperatures higher than temperature of  confinement-deconfinement phase transition we observe an increasing of CMST , and at  high enough temperatures we see that   CMST is proportional to $T^{2}$. So one can conclude that QGP  is still strongly interacted even at very high temperatures

\begin{table}
\caption{Transition temperature $T_c$ for massless quarks, $n_f=0,2,3$ (the upper part), and for different nonzero $m_q$ and $n_f$ (the lower part) in comparison with lattice data.}
\vspace{1cm}
\begin{center}
\label{tab01}
\begin{tabular}{|c|c|c|c|c|c|}
  \hline
  $n_f$ & $m_u,$ MeV & $m_d,$ MeV & $m_s,$ MeV & $T_c,$ MeV & $T_c$ (lat), MeV \\
  \hline
  0 & - & - & - & 268 & 276 \cite{LS-54}  \\
    \hline
  2 & 0 & 0 & - & 188 &  \\
    \hline
  3 & 0 & 0 & 0 & 174 &  \\
  \hline
  2 & 3 & 5 & - & 189 & 195-213 \cite{LS-55} \\
    \hline
  2+1 & 3 & 5 & 100 & 182 & 175 \cite{LS-56,LS-57} \\
    \hline
  3 & 100 & 100 & 100 & 195 & 205 \cite{LS-58}\\
  \hline
\end{tabular}
\end{center}
\end{table}

\begin{table}
\caption{Quark mass dependence of transition temperature with $|\lan \bar{q}q \ran| = (0.13$ GeV$)^3$ in comparison with the lattice data from \cite{LS-54}.}
\vspace{1cm}
\begin{center}
\label{tab02}
\begin{tabular}{|c|c|c|c|c|c|c|c|}
  \hline
  $m_q$, MeV & 25 & 50 & 100 & 200 & 400 & 600 & 1000 \\
  \hline
  $T_c$ (lat), MeV & 180 & 192 & 199 & 213 & 243 & 252 & 270 \\
    \hline
  $T_c$, MeV & 179 & 185 & 195 & 213 & 245 & 273 & 320 \\
  \hline
\end{tabular}
\end{center}
\end{table}

\begin{table}[!htb]
\caption{The temperature dependence of the quark condensate ratio $K_q(T)=\frac{\Sigma(T)}{\Sigma(0)}$ from eq.(\ref{29}) in comparison with the lattice data from \cite{LS-66}}
\begin{center}
\label{tab03}
\begin{tabular}{|l|c|c|c|c|c|c|c|c|c|c|}
\hline
$T$(in MeV) & 0 &113&122&130&142&148&153&163&176&189\\
  \hline
$K_q^{lat}(T,0)$  & 1 &0.90&0.84&0.80&0.68&0.57&0.49&0.26&
0.08&0\\
  \hline
$K_q^{th}(T,0)$ &1 & 0.85& 0.79& 0.72& 0.6& 0.51& 0.43& 0.22& 0& 0\\
\hline
\end{tabular}
\end{center}
\end{table}

\begin{table}[!htb]
\caption{The average light quark condensate as a function of temperature T in our equation (\ref{29}) vs lattice data from \cite{LS-67}.}
\begin{center}
\label{tab04}
\begin{tabular}{|l|c|c|c|c|c|}
\hline
$T(MeV)$ &135&145&155&165&175\\
  \hline
$\Sigma_{th}$&20.5&17&12.4&5.8&0\\
  \hline
$\Sigma_{lat}$&22&18&13&8&5\\
\hline
\end{tabular}
\end{center}
\end{table}

\section{Colour-electric string tension  at non-zero T}
In this secrion we will discuss colour-eletric string tension (CEST).
		In all cases the type of deconfinement transition is defined by the combination of the basic properties of the string tension
obtained from the field correlators \cite{46,47,48} -- and by the thermodynamic ensemble of the QCD matter \footnote{We will discuss in this section both QCD and gluodynamics.}. In what follows we shall discuss both fundamental microscopic and thermodynamic aspects of the deconfinement phenomenon. Here the basic role is played
by the vacuum energy density -- gluon condensate $\epsilon$ which is a fundamental concept in the dynamics of elementary particles as was suggested in \cite{49,50,51,52}.
The next important step is using of relationshp between the CEST $\sigma_E(T)$ and the gluonic condensate  $\epsilon$,  that was found in  \cite{53}.  In this way all confined dynamics is directly defined by  temperature dependence of gluonic condensate. The free energy in the confined phase $F_1(T)$ is defined as
\be
|F_1(T)|=  |\epsilon_{vac}(T)| + P_h(T), \epsilon_{vac}= 1/2 \epsilon_g + \epsilon_q, \epsilon_q= \sum_q m_q \langle \bar q q \rangle
\label{01} \ee
where the gluon condensate enters as follows
\be
\epsilon_g(T)= -\frac{b \alpha_s \langle G^2 \rangle}{32\pi},\quad \langle G^2 \rangle=\langle G^a_{\mu\nu}G^a_{\mu\nu} \rangle . \label{02} \ee
We have taken into account in the eq.(\ref{01}) that only the colorelectric gluon condensate $\langle G_E^2 \rangle= 1/2 \langle G^2 \rangle$ is connected with the pressure which is accounted for by the coefficient $1/2$ before $\epsilon_g(T)$. We also note that the quark condensate
term $\epsilon_q$ as found in \cite{41} contributes around $(10-15)$ percent of the total vacuum energy and disappears at the same temperature $T_c$ and therefore below we disregard it in the first approximation.
In eq.(\ref{01}) $P_h(T)$ is the hadron interacting gas pressure growing with the temperature while the gluon condensate $\langle G^2(T) \rangle$ (its colorelectric part) decreases, vanishing at $T=T_c$. At the same time the colormagnetic part develops independently and finally grows at large T. In this way the colorelectric and colormagnetic d.o.f. are disconnected (in the first approximation).

At this point one must define the behavior of the vacuum energy $\epsilon_g(T)$ as a function of the temperature which will
explain the properties of the hadron gas and its deconfinement transition. In what follows we impose the following condition
on the confining free energy which will be called The Vacuum Dominance Mechanism (VDM) where the hadronic pressure is growing with
temperature $T$ with the simultaneous decrease of the vacuum energy (gluon condensate), so that their sum is kept constant.
\be
|F_1(T)|= 1/2 |\epsilon(T)| + P_h(T)= 1/2|\epsilon(T=0)|. \label{03} \ee
In this way the basic QCD quantity -- the gluon condensate \cite{49} $$\langle G_E^2(0) \rangle = \dfrac{32\pi \epsilon(0)}{b\alpha_s}$$  defines the properties of the deconfining process. In particular, taking eq.(\ref{03}) at $T=T_c$  and accounting for the equality
$P_h(T_c)= P_{qgl}(T_c)$ one obtains the equation which defines
the deconfining temperature $T_c$

\be
P_q(T_c) + P_g(T_c)= 1/2|\epsilon_g(0)|.
\label{04} \ee
With such a reasoning one can calculate deconfinement temperature and can observe reasonable agreement between FCM and lattice predictions \cite{LUSI}.
The next important result that one can obtain from our assumptions is relation    between the CEST and the gluonic condensate $G_2(T)$
\be
G_2(T)= \alpha_s/\pi \langle 0|(G^a_{\mu\nu})^2|0 \rangle= \alpha_s/\pi \langle G^2 \rangle = 1.69 \sigma^2 \alpha_s^2. \label{06} \ee
At this point one takes into account that the colorelectric (CEST) $\sigma$ can be connected only to the CE part of $\langle G^2 \rangle$
\be
\sigma_E(T)= \sqrt{\frac{1/2 G_2(T)}{1.69\alpha_s^2}} \label{07} \ee
and normalizing at $T=0$  one obtains $\sigma^2(T) \approx 5.4 G_2(T)$
where we neglect the $T$ dependence of the $\alpha(T)$.
This equation allows to find the T dependence of the colourelectric gluon condensate up to its vanishing at $T=T_c$. Now due to eq.(\ref{06})
one can find the fast decreasing behavior of $\sigma_E(T)$  in the same region

\be
\sigma^2_E(T)= \sigma^2_E(0) - \frac{\xi 64 }{b} P_h(T),\,\,\xi \approx 5.4. \label{09} \ee
One can write for the noninteracting system of mesons or glueballs in the form of  Hadron Resonance Gas pressure
\be
P_h(T)= \sum_i P^i_h(T)= \frac{g_i T^2}{2\pi^2} \sum_{n=1,2,..}\frac{m_i^2 K_2(nm_i/T)}{n^2} \label{10} \ee
where $g_i$ is a multiplicity of hadrons of the type $i$ , $m_i= m_i(T)$ is the hadron mass depending on $T$ via the string tension $\sigma(T)$ and $K_2(z)$ is  $K_n(z),n=2$ the Kelvin special function.
Here one can use the following representation with the definition $z= m/T$
\be
\sum_{n=1,2,...}\frac{K_2(z)}{n^2}= \frac{1}{3z^2}\int_0^{\infty}\,dt\,\frac{t^4}{f(t,z)(\exp((f(t,z)))-1)},
\,\,f(t,z)=\sqrt{t^2+z^2} \label{11} \ee
and as a result one obtains
\be
P_h(T)=\sum_{i}\frac{g_i T^4}{6\pi^2}\Phi(z_i),\quad \Phi(z)=\int_0^{\infty}\,dt\,\frac{t^4}{f(t,z)(\exp{(f(t,z))}-1)}.\label{12} \ee

We now turn to the string tension behavior, where the normalized string tension is connected to $G_2(T)$ as  $\sigma(T)^2= 5.4 G_2(T)$ and using the relation $\epsilon(T)= b/32 G_2(T)= \frac{b \sigma^2(T)}{172.8}$

one obtains the connection between the decreasing string tension $\sigma(T)$ and the growing hadronic pressure $P_h(T)$
\be
\sigma^2(T)= c_B (P_2(T_c)- P_h(T)),\quad c_B=345/b . \label{13} \ee

 As a result using eq.(\ref{13}) one obtains for the whole region in the interval $[0,T_c]$  and  the behavior of $\sigma_E(T)$ is

\be
\sigma^2_E(T)= \sigma^2_E(0) ( 1 - P_h(T)/P_h(T_c)). \label{24} \ee

One can see in eq.(\ref{12}) that $P_h(T)=T^4 Z(M/T)$ where $Z(M/T)$ is finite for $M=0$ and therefore one can write
 for $T$ near $T_c$ neglecting the $Z'(T_c)$ contribution

\be
\sigma^2_E(T)=\sigma^2_E(0)(1-(T/T_c)^4). \label{25} \ee
One can see that the fast growth of the $P_h(T)$ near $T_c$ in eq.(\ref{25}) makes the transition curve of $\sigma(T)$ very steep for both light and heavy  hadrons. \\
As the last step  from our assumptions we can obtain  the temperature  dependence of the quark condensate.  
 It was found that  quark condensate $\Sigma(T)$ behaves in general similarly to the CE string tension as a function of
$T$ for small quark masses $m_q$ but absolute values of $\Sigma(m_q,T)$ strongly decrease for large $m_q$. In the framework of the FCM the two phenomena: the quark confinement and the quark condensate - can be directly connected since the quark condensate as the quark Green's function $S(x,x)$ at one point is considered as a closed circular-like quark trajectory with the confining surface inside.This allows to write the condensate via the quadratic Green's function $G(x,y)$ and its mass eigenvalues $M_q$ as $\Sigma_q=N_c (m_q +M_q)G(x,x)$ and expanding $G(x,x)$ in the infinite set of eigenfunctions $\phi_n$ one obtains as in \cite{61}
\be
 \Sigma_q= N_c  m_q+ M_q  \sum_n \frac{|\phi_n(0)|^2}{M_n}, \label{27}\ee

where $\phi_n(r)$ and $M_n$ are the pseudoscalar (PS) $q\bar q$ eigenfunctions and eigenvalues of the QCD Hamiltonian.
The latter are expressed via the string tension and $\alpha_s$ and neglecting the latter in the first approximation one
obtains a simple connection

\be
\Sigma_q(T)= c (\sigma_E(T))^{3/2} , \label{28} \ee

one can find the ratio $\frac{\sigma(T)}{\sigma(0)}= \eta(T)$ using (\ref{25})  which gives $\eta(T)=(1-(T/T_c)^4)^{1/2}$
and finally one obtains the deconfining behavior of the quark condensate
\be
 K_q^{th}(T)=\Sigma_q(T)/\Sigma_q(0)= \left(\frac{\sigma(T)}{\sigma(0)}\right)^{3/2}= (1- (T/T_c)^4)^{3/4} .\label{29} \ee
Numerical  results of this equation can be found in \cite{LUSI}.

\begin{figure}[!htb]
\begin{center}
\includegraphics[width=80mm,keepaspectratio=true]{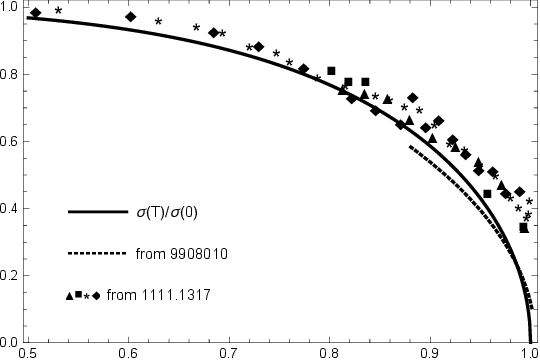}
\caption{Comparison of the lattice data for the ratio $\sigma(T)/\sigma(0)$ from \cite{LS-30} -- dotted line, with our result from eq.(\ref{25}) -- solid line, and lattice data from \cite{LS-30*} -- dots.}
\end{center}
\label{fig02}
\end{figure}

\begin{figure}[!htb]
\begin{center}
\includegraphics[width=80mm,keepaspectratio=true]{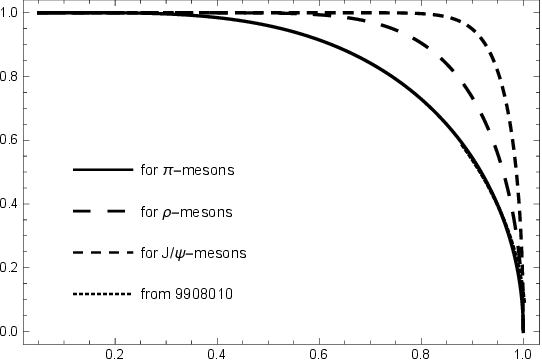}
\caption{The ratio $\sigma(T)/\sigma(0)$ according to eqs. (\ref{10}), (\ref{24}) for different values of the hadron mass: $m_{\pi}$ -- solid line, $m_{\rho}$ -- long dashed line, $m_J/\psi$ -- dashed line, and the same ratio from lattice data of \cite{LS-30} -- dotted line fully covered by the pion curve.}
\end{center}
\label{fig03}
\end{figure}

\begin{figure}[!htb]
\begin{center}
\includegraphics[width=80mm,keepaspectratio=true]{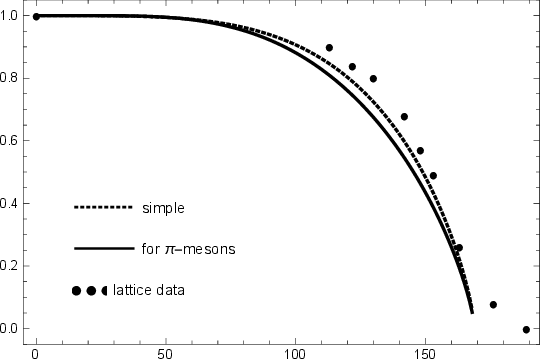}
\caption{The behavior of the quark condensate as a function of temperature T: the dotted line -- the simple form of eq. (\ref{29}), the solid line corresponds to the string tension in the pionic hadron gas to the power $4/3$, dots -- lattice data from \cite{LS-66}.}
\end{center}
\label{fig04}
\end{figure}

\section{Conclusion and Discussions}
In this brief report we discussed main properties of colour-electric and colour-magnetic string tensions and their temperature dependence. We showed that their behaviour with temperature are  quite different. If CMST is rapidly growing with temperature ,then the CEST  vanishes after some limit. We also showed that the CMST could be calculated within FCM  framework at any temperature.    Concerning the CEST we showed that there is deep interconnection between gluon condensate  and  string tension, and thus  with  some additional assumptions one can obtain the CEST  temperature dependence  from the temperature dependences of  gluon condensate and   QCD thermodynamics. As a test we tried to obtain the  quark condensate dependence from the temperature in a reasonable agreement with lattice data. Summarizing one can see, that the general construction of the theory is based on  the vacuum (gluon and quark) condensate  defining finally both colorelectric (CE) and  colormagnetic (CM) confinement. In the CE case the temporal vacuum field components are connected with the pressure which finally destroys the CE vacuum condensate and the CE string tension(deconfinement), in the CM case in the 3d dynamics the CM string tension grows with $T$.
\section{Acknowledgments}
We thanks  the Institute of high energy physics for an opportunity to discuss  all the  mentioned problems on XXXV International Workshop on High Energy Physics "From Quarks to Galaxies: Elucidating Dark Sides". 

\end{document}